\documentclass[submission,copyright,creativecommons]{eptcs}
\providecommand{\event}{Linearity 2016} 
\usepackage{breakurl}             
\usepackage{graphicx}
\usepackage{amssymb}
\usepackage{proof,pb-diagram}

\title{Linear Exponential Comonads without Symmetry}
\author{Masahito Hasegawa
\institute{Research Institute for Mathematical Sciences\\
Kyoto University\\
Kyoto, Japan}
\email{hassei@kurims.kyoto-u.ac.jp}
}
\def\titlerunning{Linear Exponential Comonads without Symmetry}
\def\authorrunning{M. Hasegawa}
\begin{document}
\maketitle

\newcommand{\comment}[1]{}
\newtheorem{theorem}{Theorem}
\newtheorem{lemma}{Lemma}
\newtheorem{proposition}{Proposition}
\newtheorem{corollary}{Corollary}
\newtheorem{definition}{Definition}
\newtheorem{remark}{Remark}
\newtheorem{example}{Example}

\newcommand{\larrow}{\rotatebox[origin=c]{180}{\,$\multimap$\,}}
\newcommand{\rarrow}{\multimap}
\newcommand{\Set}{\mathbf{Set}}
\newcommand{\val}[1]{\|#1\|}

\begin{abstract}
The notion of linear exponential comonads on symmetric monoidal categories 
has been used for modelling the exponential 
modality of linear logic. In this paper we introduce linear exponential comonads
on general (possibly non-symmetric) monoidal categories, and show some basic results on
them. 
\end{abstract}

\section{Introduction}

There are two major approaches to the categorical models of the exponential modality $!$ of 
linear logic \cite{Gir87}. 
The first is as a {\em comonad} which respects the symmetric monoidal structure 
(the multiplicative conjunction) and creates commutative comonoids (for modelling weakening and 
contraction). 
Since the pioneering work of Seely \cite{See89} and Lafont \cite{Laf88}, this direction was extensively studied by
Benton, Bierman, de Paiva and Hyland \cite{BBdePH93},  and now there is a well-accepted notion of
{\em linear exponential comonads} \cite{Bie95,HS03}.
Another one, led by Benton \cite{Ben95} and independently by Barber and Plotkin \cite{BP97},  is as an {\em adjunction} between a cartesian category (modelling non-linear proofs) and symmetric monoidal category (modelling linear proofs)
which respects the cartesian structure and symmetric monoidal structure. 
Such a situation is neatly captured as a {\em symmetric monoidal adjunction} between
a cartesian category and a symmetric monoidal category.
These two approaches are in harmony: any linear exponential comonad on
a symmetric monoidal category gives rise to a symmetric monoidal adjunction 
between a cartesian category (of coalgebras) and the symmetric monoidal category,  
while a symmetric monoidal adjunction between a cartesian category and symmetric 
monoidal category induces a linear exponential comonad on the symmetric monoidal category.
See \cite{deP14}  for a compact survey of these results, and for \cite{Mel09} 
more detaied accounts and proofs.

In this paper we consider a generalization of these categorical axiomatics for exponential 
modality to the non-symmetric setting, i.e., on monoidal categories which may not be symmetric.
This work is motivated by the desire that we should have a uniform way of modelling 
exponential modality in symmetric/non-symmetric/braided monoidal categories
(cf. the authors' ad hoc treatment of the braided case via braided monoidal comonads \cite{Has12}),
while we also hope that this work leads to a better understanding of modalities in non-commutative
(linear) logics \cite{Abr91,BG95,AR00} in general.
For the approach based on adjunctions, the answer seems more or less obvious: 
replace symmetric monoidal categories by monoidal categories, and symmetric monoidal
adjunctions by monoidal adjunctions. The corresponding axiomatics in terms of comonads
is more difficult, as the definition of linear exponential comonads heavily replies on the
presence of symmetry. 
Nevertheless, we do give an appropriate notion of linear exponential comonads without symmetry,
which however involves a number of non-trivial coherence axioms. We show that our generalization
enjoys a good correspondence with the axiomatics based on monoidal adjunctions.
After providing a few typical examples which motivate this research, we conclude this paper
with some issues for future research.

\section{Linear exponential comonads with  symmetry}

Let us recall the notion of linear exponential comonads on {\em symmetric} monoidal
categories --- for emphasizing the presence of symmetry, below we will call them
{\em symmetric} linear exponential comonads. 

\begin{remark}
We assume that the reader is familar with the notions of monoidal categories,
monoidal functors, monoidal natural transformations, monoidal comonads as well
as monoidal adjunctions \cite{EK66,Kel74}. A detailed explanation of these concepts 
aimed at the computer science/logic audience can be found in \cite{Mel09}.
Note that, in this paper, by a monoidal functor we mean a lax monoidal functor,
thus a functor $F$ equipped with possibly non-invertible coherent 
arrow $m_I:I\rightarrow FI$ and natural transformation $m_{X,Y}:FX\otimes FY\rightarrow F(X\otimes Y)$; when they are invertible, $F$ is said to be strong monoidal.
\end{remark}

\begin{definition} \cite{Bie95,HS03}
Let $\mathcal{C}$ be a  {symmetric} monoidal category.
A {\em symmetric linear exponential comonad} on $\mathcal{C}$ is
a {symmetric} monoidal comonad $(!:\mathcal{C}\rightarrow\mathcal{C},\delta_X:!X\rightarrow !!X,\varepsilon_X:!X\rightarrow X,m_{X,Y}:!X\otimes!Y\rightarrow !(X\otimes Y),m_I:I\rightarrow !I)$ on $\mathcal{C}$ equipped with
monoidal natural transformations $ d_X:!X\rightarrow !X\otimes !X$ and $ e_X:I\rightarrow !X$
such that
\begin{itemize} 
\item   $ (!X,d_X,e_X)$ forms a {commutative} comonoid,
\item  $d_X$ is a coalgebra morphism from $(!X,\delta_X)$ to $(!X\otimes !X, m_{!X,!X}\circ(\delta_X\otimes\delta_X))$,
\item  $ e_X$ is a coalgebra morphisms from $(!X,\delta_X)$ to $(I,m_I)$, and
\item  $\delta_X$ is a comonoid morphism from $(!X,d_X,e_X)$ to $(!!X,d_{!X},e_{!X})$
\end{itemize}
for each $X$.
\end{definition}

\begin{theorem}
Any symmetric monoidal adjunction 
$\mathcal{X}
\renewcommand\arraystretch{0.1}
\begin{array}{c}
\mbox{\footnotesize$F$}\\
\longrightarrow\\
\mbox{\tiny$\bot$}\\
\longleftarrow\\
\mbox{\footnotesize$U$}
\end{array}
\mathcal{C}$
between a cartesian category $\mathcal{X}$ and a symmetric monoidal category $\mathcal{C}$
gives rise to a symmetric linear exponential comonad $FU$ on $\mathcal{C}$.
\end{theorem}
\underline{Proof.} Follows from routine calculation. 
$\square$

\begin{theorem}
Given a symmetric linear exponential comonad $ !$ on a
{symmetric} monoidal category $ \mathcal{C}$,
its category of Eilenberg-Moore coalgebras $ \mathcal{C}^!$ is a cartesian category, with
$$\!\!\!\!\!\!\!\!
\begin{array}{rcl}
(A,\alpha:A\rightarrow !A)\times(B,\beta:B\rightarrow !B)
&=&
(A\otimes B,A\otimes B\stackrel{\alpha\otimes\beta}{\longrightarrow}!A\otimes !B\stackrel{m_{A,B}}{\longrightarrow}!(A\otimes B))\\
1 &=& (I,I\stackrel{m_I}{\longrightarrow}!I)
\end{array}
$$\noindent
Moreover, the comonadic adjunction 
$\mathcal{C}^!
\renewcommand\arraystretch{0.1}
\begin{array}{c}
\longrightarrow\\
\mbox{\tiny$\bot$}\\
\longleftarrow
\end{array}
\mathcal{C}$
is 
{symmetric} monoidal
with respect to the cartesian products of $\mathcal{C}^!$ and monoidal products of $\mathcal{C}$.
\end{theorem}
\underline{Proof.}
A detailed and accessible proof can be found in \cite{Mel09}. Here we shall sketch its outline.
Since $!$ is a {symmetric} monoidal comonad, the category of coalgebras $\mathcal{C}^!$ is
{symmetric} monoidal with $I=(I,m_I)$ and
$$
(A,\alpha)\otimes(B,\beta)=
(A\otimes B,A\otimes B\stackrel{\alpha\otimes\beta}{\longrightarrow}!A\otimes !B\stackrel{m_{A,B}}{\longrightarrow}!(A\otimes B)).
$$
Then we can show that
$ d_X$ and $ e_X$ form a  comonoid on the cofree coalgebra 
$ (!X,\delta_X:!X\rightarrow!!X)$
in $\mathcal{C}^!$.
Moreover, they induce a comonoid on any coalgebra $(A,\alpha)$ via the
retraction $\alpha:A\rightarrow!A$ and $\varepsilon_A:!A\rightarrow A$.
The induced comonoid structure on each coalgebra extends to natural transformations
$ d_{(A,\alpha)}:(A,\alpha)\rightarrow(A,\alpha)\otimes(A,\alpha)$ and
$ e_{(A,\alpha)}:(A,\alpha)\rightarrow(I,m_I)$.
Finally we can appeal to the folklore result that
any {symmetric} monoidal category with a natural comonoid structure on every object is
a cartesian category. 
The comonadic adjunction is symmetric monoidal because its left adjoint is a strong symmetric
monoidal functor \cite{Kel74}.
$\square$

\mbox{}

Symmetry appears in many places in the definition of symmetric linear exponential comonads
as well as the proof of Theorem 2
(sometimes rather implicitly, for instance for making $X\mapsto !X\otimes !X$ a monoidal functor). 
However, note that,  only the symmetry of the form
$ !X\otimes !Y \rightarrow !Y\otimes !X$
is needed in the proof. Moreover, such a symmetry $ !X\otimes !Y \rightarrow !Y\otimes !X$
can be re-defined from other constructs:
\begin{proposition}
In a symmetric monoidal category with a symmetric linear exponential comonad,
the morphism
$$
\dgARROWLENGTH=4em
\begin{diagram}
\node{!X\otimes\,!Y}
 \arrow{s,l}{\delta_X\otimes\delta_Y}
\node[3]{!Y\otimes\,!X}
\\
\node{!!X\otimes\,!!Y}
 \arrow{e,b}{m_{!X,!Y}}
\node{!(!X\otimes\,!Y)}
 \arrow{e,b}{d_{!X\otimes\,!Y}}
\node{!(!X\otimes\,!Y)\otimes\,!(!X\otimes\,!Y)}
 \arrow{e,b}{!(e_X\otimes id)\otimes\,!(id\otimes e_Y)}
\node{!!Y\otimes\,!!X}
 \arrow{n,r}{\varepsilon_{!Y}\otimes\varepsilon_{!X}}
\end{diagram}
$$
agrees with the symmetry $!X\otimes !Y \rightarrow !Y\otimes !X$.
\end{proposition}
A proof is given in Appendix A. 
This observation suggests that it should be possible to define a linear exponential comonad without assuming symmetry
(i.e. on any monoidal category),
by using this re-defined ``symmetry'', so that its category of coalgebras is cartesian and
the induced comonadic adjunction becomes monoidal. In the next section,
we give such a definition of linear exponential comonads on monoidal categories.

\begin{remark}
A proof corresponding to this
re-defined ``symmetry'' in multiplicative linear logic would be:
$$
\infer[Cut]{!X\otimes!Y\vdash !Y\otimes!X}
       {\infer[L\otimes]{!X\otimes !Y\vdash !(!X\otimes !Y)}
           {\infer[Promotion]{!X,!Y\vdash !(!X\otimes !Y)}
                {\infer[R\otimes]{!X,!Y\vdash !X\otimes !Y}
                     {\infer[Axiom]{!X\vdash !X}{} & \infer[Axiom]{!Y\vdash !Y}{}}}}
&
\infer[Contraction]{!(!X\otimes !Y)\vdash !Y\otimes !X}
               {\infer[R\otimes]{!(!X\otimes !Y),!(!X\otimes !Y)\vdash !Y\otimes !X}
                       {\infer[Dereliction]{!(!X\otimes !Y)\vdash !Y}
                               {\infer[L\otimes ]{!X\otimes !Y\vdash !Y}
                                       {\infer[Weakening]{!X,!Y\vdash !Y}
                                               {\infer[Axiom]{!Y\vdash !Y}{}}}}
                        &
                       \infer[Dereliction]{!(!X\otimes !Y)\vdash !Y}
                               {\infer[L\otimes]{!X\otimes !Y\vdash !X}
                                       {\infer[Weakening]{!X,!Y\vdash !X}
                                               {\infer[Axiom]{!X\vdash !X}{}}}}}}}
$$
It is not possible to remove the cut in this proof in the non-commutative setting, unless 
we assume the exchange rule on $!$-formulas. Thus our choice of
taking this derived ``symmetry'' as a basic notion, while makes a good sense
at the semantic level,  does not lead to a pleasant proof system (with the
cut-elimination property) at the syntactic level.
\end{remark}

\section{Linear exponential comonads without symmetry}

\begin{definition}
A {\em linear exponential comonad} on a monoidal category $\mathcal{C}$ 
is a 
monoidal comonad\linebreak
$ (!,\delta,\varepsilon,m,m_I)$
on $\mathcal{C}$
equipped with  
a monoidal natural transformation
$e_X:!X\rightarrow I$ and 
a natural transformation
$d_X:!X\rightarrow !X\otimes\,!X$
such that, with 
$\sigma_{X,Y}=
(\varepsilon_{!Y}\otimes\varepsilon_{!X})\circ
(!(e_X\otimes id)\otimes\,!(id\otimes e_Y))\circ
d_{!X\otimes\,!Y}\circ
m_{!X,!Y}\circ
(\delta_X\otimes\delta_Y):
!X\otimes !Y\rightarrow !Y\otimes !X
$,
\comment{
$\sigma_{X,Y}:!X\otimes\,!Y\rightarrow !Y\otimes\,!X$
defined by
$$
\dgARROWLENGTH=4em
\begin{diagram}
\node{!X\otimes\,!Y}
 \arrow{s,l}{\delta_X\otimes\delta_Y}
\node[3]{!Y\otimes\,!X}
\\
\node{!!X\otimes\,!!Y}
 \arrow{e,b}{m_{!X,!Y}}
\node{!(!X\otimes\,!Y)}
 \arrow{e,b}{d_{!X\otimes\,!Y}}
\node{!(!X\otimes\,!Y)\otimes\,!(!X\otimes\,!Y)}
 \arrow{e,b}{!(e_X\otimes id)\otimes\,!(id\otimes e_Y)}
\node{!!Y\otimes\,!!X}
 \arrow{n,r}{\varepsilon_{!Y}\otimes\varepsilon_{!X}}
\end{diagram}
$$
}
\begin{enumerate}
\item[(1)]
\comment{%
the functor $ FX=!X\otimes\,!X$ is monoidal with
$ I\stackrel{m_I\otimes m_I}{\longrightarrow}!I\otimes\,!I=F(I)$ and
{
$$
FX\otimes FY=!X\otimes\,!X\otimes\,!Y\otimes\,!Y
 \stackrel{id\otimes\sigma\otimes id}{\longrightarrow}
!X\otimes\,!Y\otimes\,!X\otimes\,!Y 
 \stackrel{m\otimes m}{\longrightarrow}
!(X\otimes Y)\otimes\,!(X\otimes Y)=F(X\otimes Y)
$$}\noindent
which means that 
}%
the following diagram commutes:
{
$$\!\!\!\!\!\!\!\!\!\!\!\!\!\!\!\!\!\!\!\!\!\!\!\!
\begin{diagram}
\dgARROWLENGTH=1.5em
\node{!X\otimes\,!X\otimes\,!Y\otimes\,!Y\otimes\,!Z\otimes\,!Z}
 \arrow{s,l}{id\otimes\sigma\otimes id}
 \arrow{e,t}{id\otimes\sigma\otimes id}
\node{!X\otimes\,!Y\otimes\,!X\otimes\,!Y\otimes\,!Z\otimes\,!Z}
 \arrow{e,t}{m\otimes m\otimes id}
\node{!(X\otimes Y)\otimes\,!(X\otimes Y)\otimes\,!Z\otimes\,!Z}
 \arrow{s,r}{id\otimes\sigma\otimes id}
\\
\node{!X\otimes\,!X\otimes\,!Y\otimes\,!Z\otimes\,!Y\otimes\,!Z}
 \arrow{s,l}{id\otimes m\otimes m}
\node[2]{!(X\otimes Y)\otimes\,!Z\otimes\,!(X\otimes Y)\otimes\,!Z}
 \arrow{s,r}{m\otimes m}
\\
\node{!X\otimes\,!X\otimes\,!(Y\otimes Z)\otimes\,!(Y\otimes Z)}
 \arrow{e,b}{id\otimes\sigma\otimes id}
\node{!X\otimes\,!(Y\otimes Z)\otimes\,!X\otimes\,!(Y\otimes Z)}
 \arrow{e,b}{m\otimes m}
\node{!(X\otimes Y\otimes Z)\otimes\,!(X\otimes Y\otimes Z)}
\end{diagram}
$$}
\item[(2)]
$ m_{!Y,!X}\circ\sigma_{!X,!Y} = !\sigma_{X,Y}\circ m_{!X,!Y}$, 
\item[(3)]
$\sigma^{-1}_{X,Y}=\sigma_{Y,X}$,
\item[(4)] the following diagram commutes: 
{
$$
\begin{diagram}
\node{!X\otimes\,!Y\otimes\,!Z}
 \arrow{s,l}{id\otimes\sigma_{Y,Z}}
 \arrow{e,t}{\delta_X\otimes\delta_Y\otimes id}
\node{!!X\otimes\,!!Y\otimes\,!Z}
 \arrow{e,t}{m_{!X,!Y}\otimes id}
\node{!(!X\otimes\,!Y)\otimes\,!Z}
 \arrow{e,t}{\sigma_{!X\otimes\,!Y,Z}}
\node{!Z\otimes\,!(!X\otimes\,!Y)}
 \arrow{s,r}{id\otimes\varepsilon_{!X\otimes\,!Y}}\\
\node{!X\otimes\,!Z \otimes\,!Y}
 \arrow[3]{e,b}{\sigma_{X,Z}\otimes id}
\node[3]{!Z\otimes\,!X\otimes\, !Y}
\end{diagram}
$$}
\item[(5)]
the following diagrams commute,
{
\dgARROWLENGTH=1.5em
$$\!\!\!\!\!\!\!\!\!\!\!\!\!\!\!\!\!\!\!\!\!\!\!\!
\begin{diagram}
\node{!X\otimes\,!Y}
 \arrow{s,l}{m}
 \arrow{e,t}{d\otimes d}
\node{!X\otimes\,!X\otimes\,!Y\otimes\,!Y}
 \arrow{e,t}{id\otimes\sigma\otimes id}
\node{!X\otimes\,!Y\otimes\,!X\otimes\,!Y}
 \arrow{s,r}{m\otimes m}
\node{I}
 \arrow{s,l}{m_I}
 \arrow{se,t}{m_I\otimes m_I}
\\
\node{!(X\otimes Y)}
 \arrow[2]{e,b}{d}
\node[2]{!(X\otimes Y)\otimes\,!(X\otimes Y)}
\node{!I}
 \arrow{e,b}{d}
\node{!I\otimes\,!I}
\end{diagram}
$$}
\item[(6)] 
$ (!X,e_X,d_X)$ is a comonoid in $\mathcal{C}$,
\item[(7)]
$e_X$ and $d_X$ are coalgebra morphisms, and
\item [(8)]
$\delta_X$ is a comonoid morphism.
\end{enumerate}
\end{definition}

\begin{remark}
In this definition, the conditions (1)-(5) involve $\sigma$.
Other conditions are independent of $\sigma$, and are actually the
same as the conditions for the symmetric linear exponential comonads.
\end{remark}

\begin{theorem}
Any monoidal adjunction 
$\mathcal{X}
\renewcommand\arraystretch{0.1}
\begin{array}{c}
\mbox{\footnotesize$F$}\\
\longrightarrow\\
\mbox{\tiny$\bot$}\\
\longleftarrow\\
\mbox{\footnotesize$U$}
\end{array}
\mathcal{C}$
between a cartesian category $\mathcal{X}$ and a monoidal category $\mathcal{C}$
gives rise to a linear exponential comonad $FU$ on $\mathcal{C}$.
\end{theorem}
\underline{Proof:}
Follows from routine calculation.
$\square$

\begin{theorem}
Given a linear exponential comonad $!$ on a  monoidal category $\mathcal{C}$,
its category of coalgebras $\mathcal{C}^!$ is cartesian.
The comonadic adjunction between $\mathcal{C}^!$ and $\mathcal{C}$ is monoidal.
\end{theorem}
\underline{Proof:}
The proof is analogous to the symmetric case, with some extra care on the use of
$\sigma$ instead of symmetry. (In fact, our conditions on linear exponential comonads
are chosen so that the proof can mimick the symmetric case.)
The condition (1) implies that 
the functor $\Delta X=!X\otimes\,!X$ is monoidal with
$ I\stackrel{m_I\otimes m_I}{\longrightarrow}!I\otimes\,!I=\Delta(I)$ and
$$
\Delta X\otimes \Delta Y=!X\otimes\,!X\otimes\,!Y\otimes\,!Y
 \stackrel{id\otimes\sigma\otimes id}{\longrightarrow}
!X\otimes\,!Y\otimes\,!X\otimes\,!Y 
 \stackrel{m\otimes m}{\longrightarrow}
!(X\otimes Y)\otimes\,!(X\otimes Y)=\Delta(X\otimes Y).
$$
The condition (2) is for making ! ``symmetric monoidal'' with respect to $\sigma$.
The conditions (3) and (4) imply that $\sigma$ behaves like a symmetry.
The condition (5) implies that $d$ is a monoidal natural transformation from $!$ to $\Delta$
defined as above. The condition (6), together with the fact that the (co-)commutativity
$\sigma_{X,X}\circ d_X=d_X$ is derivable from other axioms, says that 
$(!X,d_X,e_X)$ is a ``commutative'' comonoid.
Then we are able to mimick the proof for the case with symmetry. First, we can show that
$\mathcal{C}^!$ is symmetric monoidal --- $\sigma$ extends to a symmetry on $\mathcal{C}^!$.
Second, we have that $d_X$ and $e_X$ from a comonoid on the cofree coalgebra on $X$,
which naturally extends to all coalgebras. Thus $\mathcal{C}^!$ is a symmetric monoidal
category with a natural comonoid structure on all objects.
$\square$

\mbox{}

A symmetric linear exponential comonad can be characterized 
as a symmetric monoidal comonad such that the induced symmetric monoidal structure
on the category of coalgebras is cartesian \cite{Man04}. 
As a corollary to the theorems above, this characterization  
extends to the non-symmetric case, just by dropping all ``symmetric'': 

\begin{theorem}
A monoidal comonad is a linear exponential comonad if and only if
the induced monoidal structure on the category of coalgebras is cartesian. 
\end{theorem}

The following results, which are standard in the symmetric case, 
easily follow from the theorems above.
\begin{proposition}
In a monoidal category $\mathcal{C}$ with a linear exponential comonad $!$ and finite products,
there is a natural isomorphism
 $!X\otimes !Y\cong !(X\times Y)$ as well as an isomorphism $I\cong !1$
making $!$ a strong monoidal functor from $(\mathcal{C},\times,1)$ to $(\mathcal{C},\otimes,I)$.
\end{proposition}

\begin{proposition}
Suppose that $\mathcal{C}$ is a monoidal (left or right) closed category with a
linear exponential comonad $!$. Then there exists a cartesian closed category
$\mathcal{X}$ and a monoidal adjunction
$\mathcal{X}
\renewcommand\arraystretch{0.1}
\begin{array}{c}
\mbox{\footnotesize$F$}\\
\longrightarrow\\
\mbox{\tiny$\bot$}\\
\longleftarrow\\
\mbox{\footnotesize$U$}
\end{array}
\mathcal{C}$
such that $!$ agrees with the induced comonad $FU$. In addition, if $\mathcal{C}$
has finite products, the co-Kleisli category $\mathcal{C}_!$ is cartesian closed,
and the co-Kleisli adjunction is monoidal.
\end{proposition}
Note that, in a monoidal bi-closed category (with $X\otimes(\_)\dashv X\rarrow(\_)$ and $(\_)\otimes X\dashv (\_)\larrow X$) with a linear exponential comonad $!$,
$!X\rarrow Y$ may not be isomorphic to $Y\larrow !X$, but they are isomorphic in 
the co-Kleisli category.

\section{Examples}

Obviously, symmetric linear exponential comonads are instances of 
our linear exponential comonads:

\begin{proposition}
A linear exponential comonad on a symmetric monoidal category is a symmetric linear exponential comonad
if and only if  $\sigma$ (given in Definition 2)  agrees with the symmetry.
\end{proposition}

\comment{
For a simple posetal example invalidating $!X\otimes Y=Y\otimes !X$:
\begin{example}
Fix a monoid $(M,\cdot,e)$ and consider the poset $(2^M,\leq)$
where $X\leq Y$ iff $Y\subseteq X$. 
With $I=\{e\}$ and $X\otimes Y=\{xy~|~x\in X,y\in Y\}$,
$(2^M,\leq)$ becomes a monoidal poset. By letting
$!X=M$ for any $X$, we have a linear exponential comonad $!$ on $(2^M,\leq)$.
\end{example}
}

Here is a simple example of a non-symmetric monoidal bi-closed category
with a linear exponential comonad.
\begin{example}
Let $M=(M,\cdot,e)$ be a (non-commutative) monoid and consider the slice category $\Set/M$.
Thus an object of $\Set/M$ is a set $A$ equipped with
an $M$-valued map $\val{\_}_A:A\rightarrow M$ (often the subscript will be omitted),
and a morphism $f:(A,\val{\_}_A)\rightarrow(B,\val{\_}_B)$ in $\Set/M$ 
is a map $f:A\rightarrow B$ such that $\val{f(a)}=\val{a}$ holds for any $a\in A$.
$\Set/M$ is monoidal bi-closed with
$$
\begin{array}{rcl}
I &=& \{*\}~\mbox{with}~\val{*}=e\\
A\otimes B &=& A\times B~\mbox{with}~\val{(a,b)}=\val{a}\cdot\val{b}\\
B\larrow A &=& 
 \{(x,f)~|~x\in M, f:A\rightarrow B~\mbox{s.t.}~\val{f(a)}=x\cdot\val{a}~\mbox{for}~a\in A\}~\mbox{with}~\val{(x,f)}=x\\
A\rarrow B &=& 
 \{(x,f)~|~x\in M, f:A\rightarrow B~\mbox{s.t.}~\val{f(a)}=\val{a}\cdot x~\mbox{for}~a\in A\}~\mbox{with}~\val{(x,f)}=x\\
\end{array}
$$
There is a linear exponential comonad $!$ on $\Set/M$ 
given  by  $!A=\{a\in A~|~\val{a}=e\}$ with $\val{a}=e$, 
whose category of coalgebras is equivalent to $\Set$.
\comment{
(Via the equivalence $\Set/M\cong \Set^M$, it is not hard to see
that this monoidal bi-closed structure is equivalent to that on $\Set^M$ with 
Day's tensor product, where $M$ is regarded as a discrete strict monoidal category.)}
\end{example}

\comment{
Here is an example of non-symmetric monoidal bi-closed categories
which is a simple instance of Day's tensor product on presheaves \cite{Day70}.

\begin{example}
Let $\mathcal{C}$ be a cartesian closed category
with finite coproducts (e.g. $\mathbf{Set}$).
There is a monoidal bi-closed structure on $\mathcal{C}^3$ which is not
symmetric unless $\mathcal{C}$ is trivial. Define:
$$
\begin{array}{rcl}
I &=& (1,0,0)\\
(X_0,X_1,X_2)\otimes(Y_0,Y_1,Y_2) &=& \displaystyle
(X_0\times Y_0,~
 \sum_{i=0}^2 X_i\times Y_1 + X_1\times Y_0,~
 \sum_{i=0}^2 X_i\times Y_2 + X_2\times Y_0)\\
(X_0,X_1,X_2)\rarrow(Y_0,Y_1,Y_2) &=& \displaystyle
(\prod_{i=0}^2 Y_i^{X_i},~\prod_{i=0}^2 Y_1^{X_i},~\prod_{i=0}^2 Y_2^{X_i})\\
(Y_0,Y_1,Y_2)\larrow(X_0,X_1,X_2) &=& \displaystyle
(Y_0^{X_0}\times U,~
 Y_1^{X_0}\times U,~
 Y_2^{X_0}\times U)~~\mathrm{where}~U=Y_1^{X_1}\times Y_2^{X_2}
\end{array}
$$
Then we have 
$
\mathcal{C}^3(\mathbf{Y},\mathbf{X}\rarrow\mathbf{Z})
\cong
\mathcal{C}^3(\mathbf{X}\otimes\mathbf{Y},\mathbf{Z})
\cong
\mathcal{C}^3(\mathbf{X},\mathbf{Z}\larrow\mathbf{Y})
$.
On the other hand, we have 
$$
(0,1,0)\otimes(0,0,1) = (0,0,1)
~~~~\mbox{and}~~~~
(0,0,1)\otimes(0,1,0) = (0,1,0)
$$
which are not isomorphic unless $\mathcal{C}$ is a trivial category.
There is a linear exponential comonad $ !$ on $ \mathcal{C}^3$ with 
$!(X_0,X_1,X_2)=(X_0,0,0)$,
whose category of coalgebras is equivalent to the cartesian closed category $\mathcal{C}$.
\end{example}
}

Presheaves are a rich source of examples:
\begin{example}
Let $F:\mathcal{X}\rightarrow\mathcal{C}$ be a strong monoidal 
functor from a cartesian category $\mathcal{X}$ to a monoidal category $\mathcal{C}$.
The left Kan extension along 
$F^\mathit{op}:\mathcal{X}^\mathit{op}\rightarrow\mathcal{C}^\mathit{op}$
gives a monoidal adjunction
$$\Set^{\mathcal{X}^\mathit{op}}
\renewcommand\arraystretch{0.1}
\begin{array}{c}
\mbox{\footnotesize$\mathbf{Lan}_{F^\mathit{op}}(\_)$}\\
\longrightarrow\\
\mbox{\tiny$\bot$}\\
\longleftarrow\\
\mbox{\footnotesize$(\_)\circ F^\mathit{op}$}
\end{array}
\Set^{\mathcal{C}^\mathit{op}}$$
between the cartesian closed category $\Set^{\mathcal{X}^\mathit{op}}$
and the monoidal bi-closed category $\Set^{\mathcal{C}^\mathit{op}}$
(with the monoidal structure on $\Set^{\mathcal{C}^\mathit{op}}$
given by Day's tensor product \cite{Day70}).
From this we obtain a linear exponential comonad $!$ on $\Set^{\mathcal{C}^\mathit{op}}$
where $\displaystyle !G=\mathbf{Lan}_{F^\mathit{op}}(G\circ F^\mathit{op})=
\int^{X\in\mathcal{X}}\mathcal{C}(\_,FX)\times G(FX)$ for $G:\mathcal{C}^\mathit{op}\rightarrow\Set$.
\end{example}

\begin{remark}
The two examples above can be extended to more involved ones using
categorical glueing (given by comma categories, or
more generally change-of-base of monoidal closed bi-fibrations
along monoidal functors). Such glueing constructions are known
for the symmetric cases \cite{Has99a,Has99b,HS03}, and
can be generalized to the non-symmetric cases without much difficulty. 
\end{remark}

We conclude this section with an example 
with non-symmetric braiding \cite{JS93} taken from our previous work \cite{Has12}.

\begin{example}
Let $G$ be a group with the unit element $e$. A crossed $G$-set is a 
set $X$ equipped with a group action $\cdot:G\times X\rightarrow X$ and a map
$|\_|:X\rightarrow G$ such that, for any $g\in G$ and $x\in X$, 
$|g\cdot x|=g|x|g^{-1}$ holds.
Now let $\mathbf{XRel}(G)$ be the category whose objects are crossed $G$-sets, and 
a morphism from $(X,\cdot,|\_|)$ to $(Y,\cdot,|\_|)$ is a binary relation $r:X\rightarrow Y$
such that $(x,y)\in r$ implies $(g\cdot x,g\cdot y)\in r$ as well as $|x|=|y|$.
The composition and identity are just those of binary relations. 
$\mathbf{XRel}(G)$ is monoidal, with 
$(X,\cdot,|\_|)\otimes(Y,\cdot,|\_|)=(X\times Y,(g,(x,y))\mapsto(g\cdot x,g\cdot y), (x,y)\mapsto |x||y|)$.
This monoidal structure is not symmetric but {\em braided}:
the braiding on $(X,\cdot,|\_|)$ and $(Y,\cdot,|\_|)$ is given by
$$
\{((x,y),(|x|\cdot y,x)~|~x\in X,y\in Y\}:(X,\cdot,|\_|)\otimes(Y,\cdot,|\_|)\rightarrow(Y,\cdot,|\_|)\otimes(X,\cdot,|\_|).
$$
(In fact, $\mathbf{XRel}(G)$ forms a ribbon category \cite{Shu94,Tur94}, thus allows interpretation of tangles \cite{Has12}.)
There is a linear exponential comonad on $\mathbf{XRel}(G)$ which sends
$(X,\cdot,|\_|)$ to the finite multiset of  $\{x\in X~|~|x|=e\}/\sim$ 
(where $x\sim y$ if $g\cdot x=y$ for some $g\in G$) with trivial action
$g\cdot u=u$ and valuation $|u|=e$.
\end{example}
Note that  linear exponential comonads on braided monoidal categories 
in \cite{Has12} are the linear exponential
comonads in this paper such that $\sigma$ agrees with the braiding, thus the situation
is exactly the parallel of the symmetric case.

\section{Conclusion and future work}

We have given the notion of linear exponential comonads on
arbitrary monoidal categories, and shown that it has a good correspondence to
the axiomatics based on monoidal adjunctions. There are a number of immediate
issues which are left as the future work:

\paragraph{Simpler axiomatization}
Unfortunately, our definition of linear exponential comonads involve a number of
coherence axioms which are rather hard to be used in practice. We do expect that
some of the axioms are actually redundant, and some (conceptually or technically)
simpler axiomatization can be found. As an easy direction, it would be interesting
to consider the case with finite products (additive products) which would allow 
some substantially simpler axiomatics where the key isomorphism $\sigma_{X,Y}:
!X\otimes !Y\rightarrow !Y\otimes !X$ can be simply given by 
$!X\otimes !Y\cong !(X\times Y)\cong !(Y\times X)\cong !Y\otimes !X$.

\paragraph{Proof systems and term calculi}
The absence of symmetry causes a number of troubles at the level of syntax, i.e.,
on proof systems or term calculi (linear lambda calculi). From the proof theoretical point
of view, the nasty issue on cut-elimination must be remedied by introducing 
Exchange rules for $!$-types, whose precise formulation can be of some interest.
As a direction closer to the categorical models, it would be nice if we have a
(equationally) sound and complete term calculus for monoidal bi-closed categories
with a linear exponential comonad, like the DILL calculus \cite{BP97} for the symmetric case.
The case of (non-symmetric) $*$-autonomous categories \cite{Bar95,BD11} with a linear exponential comonad
should be also interesting, as there can be a term calculus just with linear/non-linear 
implications and the falsity type as type constructs \cite{Has05}. 

\paragraph{Issues on (2- or bi-)categories of models}
While linear exponential comonads and monoidal adjunctions are closely related, 
the (2- or bi-)categories of these structures are not (bi-)equivalent. We believe 
that the situation is the same as the symmetric case \cite{MMdePR05,deP14}, but the precise
details (the correct formulation of morphisms in particular) are yet to be examined.

\paragraph{Non-monoidal exponential modality}
The exponential modality of the non-commutative propositional linear logic in 
\cite{BG95} 
does not satisfy
the monoidality $!X\otimes !Y\rightarrow !(X\otimes Y)$ (hence promotion is not valid in general); 
some justification for 
not allowing promotion/monoidality is given in {\em ibid.}
Clearly our linear exponential comonads are not appropriate for modelling such a non-monoidal case,
while we do not know a suitable categorical axiomatics for the non-monoidal exponential modality.

\subsection*{Acknowledgements}
I thank Shin-ya Katsumata for helpful discussions, and Kazushige Terui for
his advice on non-commutative linear logic and cut-elimination.
This research is partly supported by the Grant-in-Aid for Scientific Research (C) 15K00013.

\bibliographystyle{eptcs}
\bibliography{generic}

\newpage
\appendix

\section{Proof of Proposition 1}

\comment{
\begin{proposition}
In a symmetric monoidal category with a symmetric linear exponential comonad,
the morphism
$$
\dgARROWLENGTH=4em
\begin{diagram}
\node{!X\otimes\,!Y}
 \arrow{s,l}{\delta_X\otimes\delta_Y}
\node[3]{!Y\otimes\,!X}
\\
\node{!!X\otimes\,!!Y}
 \arrow{e,b}{m_{!X,!Y}}
\node{!(!X\otimes\,!Y)}
 \arrow{e,b}{d_{!X\otimes\,!Y}}
\node{!(!X\otimes\,!Y)\otimes\,!(!X\otimes\,!Y)}
 \arrow{e,b}{!(e_X\otimes id)\otimes\,!(id\otimes e_Y)}
\node{!!Y\otimes\,!!X}
 \arrow{n,r}{\varepsilon_{!Y}\otimes\varepsilon_{!X}}
\end{diagram}
$$
agrees with the symmetry $!X\otimes !Y \rightarrow !Y\otimes !X$.
\end{proposition}
}
Let $c$ be the symmetry. The following commutative diagram
shows that
$$c_{!Y,!X}\circ(\varepsilon_Y\otimes\varepsilon_X)\circ
(!(e_X\otimes id)\otimes!(id\otimes e_Y))\circ d_{!X\otimes !Y}\circ
m_{!X,!Y}\circ(\delta_X\otimes\delta_Y)
~=~id_{!X\otimes!Y}$$
holds.

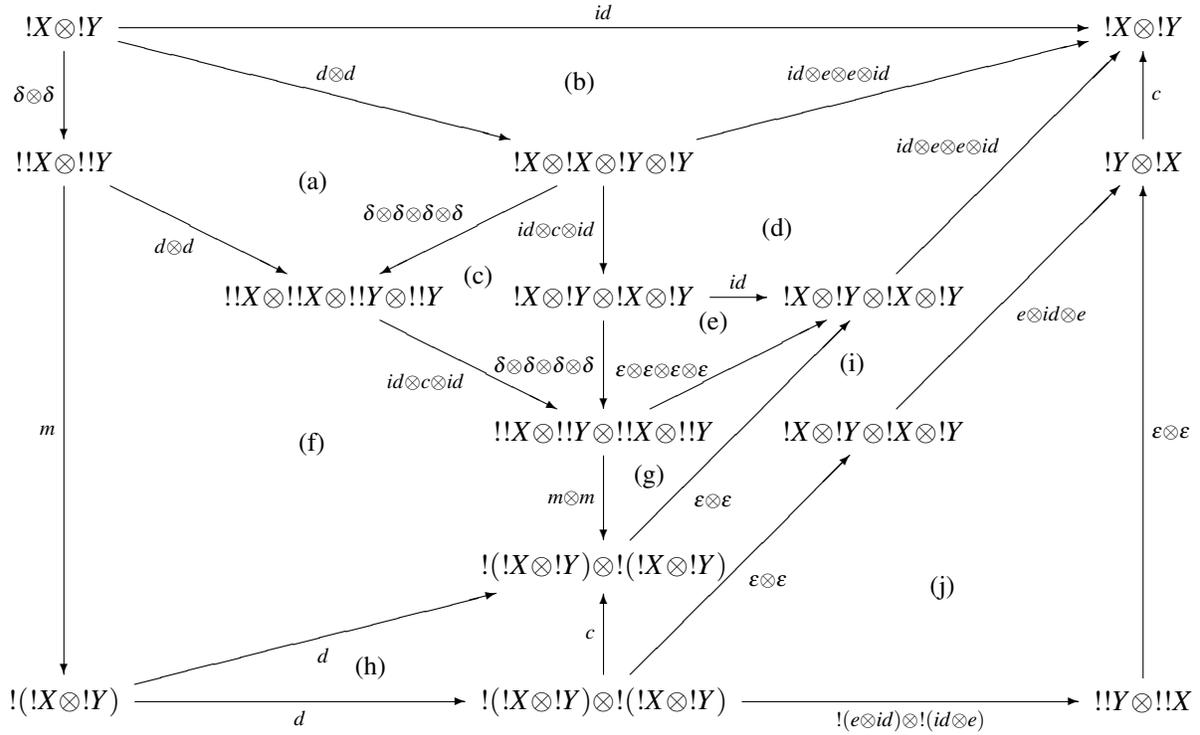
\newcommand{\ten}{\otimes}
\newcommand{\mb}[1]{\mbox{\small #1}}
$$
\dgARROWLENGTH=1.3em
\begin{diagram}
\node{!X\ten!Y}
 \arrow{s,l}{\delta\ten\delta}
 \arrow{ese,t}{d\ten d}
 \arrow[4]{e,t}{id}
\node[4]{!X\ten!Y}
\\ 
\node{!!X\ten!!Y}
 \arrow[4]{s,l}{m}
 \arrow{se,b}{d\ten d}
 \arrow[2]{e,b,!}{\mb{(a)}} 
\node[2]{!X\ten!X\ten!Y\ten!Y}
 \arrow{sw,l}{\delta\ten\delta\ten\delta\ten\delta}
 \arrow{s,t}{id\ten c\ten id}
 \arrow{ene,t}{id\ten e\ten e\ten id}
 \arrow{n,t,!}{\mb{(b)}}
\node[2]{!Y\ten!X}
 \arrow{n,r}{c}
\\ 
\node[2]{!!X\ten!!X\ten!!Y\ten!!Y}
 \arrow{se,b}{id\ten c\ten id}
 \arrow{e,t,!}{\mb{(c)}} 
 \arrow[2]{s,t,!}{\mb{(f)}} 
\node{!X\ten!Y\ten!X\ten!Y}
 \arrow{s,l}{\delta\ten\delta\ten\delta\ten\delta}
 \arrow{e,t}{id}
 \arrow{ene,t,1,!}{\mb{(d)}} 
 \arrow{se,t,1,!}{\mb{(e)}} 
\node{!X\ten!Y\ten!X\ten!Y}
 \arrow{nne,t}{id\ten e\ten e\ten id}
\\ 
\node[3]{!!X\ten!!Y\ten!!X\ten!!Y}
 \arrow{s,l}{m\ten m}
 \arrow{ne,t,1}{\varepsilon\ten\varepsilon\ten\varepsilon\ten\varepsilon\!\!\!\!\!\!\!\!}
 \arrow{s,r,1,!}{~~~\mb{(g)}} 
\node{!X\otimes!Y\otimes!X\otimes!Y}
 \arrow{nne,b}{e\otimes id\otimes e}
 \arrow{n,t,!}{\mb{(i)}}
\\ 
\node[3]{!(!X\ten!Y)\ten!(!X\ten!Y)}
 \arrow{nne,b,1}{\varepsilon\ten\varepsilon}
 \arrow[2]{e,b,!}{\mb{(j)}}
\\ 
\node{!(!X\ten!Y)}
 \arrow{ene,b}{d}
 \arrow[2]{e,b}{d}
\node[2]{!(!X\ten!Y)\ten!(!X\ten!Y)}
 \arrow{n,l}{ c}
 \arrow{nne,b}{\varepsilon\ten\varepsilon}
 \arrow[2]{e,b}{!(e\otimes id)\otimes!(id\otimes e)}
 \arrow{wnw,b,1,!}{\mb{(h)}}
\node[2]{!!Y\ten !!X}
 \arrow[4]{n,r}{\varepsilon\otimes\varepsilon}
\end{diagram}
$$

\mbox{}

\begin{center}
\begin{tabular}{ll}
(a) & $\delta$ is a comonoid morphism \\
(b) & co-unit law of  the comonoid \\ 
(c), (d), (i) & naturality of $c$ \\
(e) & co-unit law of the comonad\\
(f) & $d$ is a coalgebra morphism\\
(g) & monoidality of $\varepsilon$ \\
(h) & commutativity of the comonoid \\
(j) & naturality of $\varepsilon$
\end{tabular}
\end{center}

\mbox{}

\mbox{}\\
From this, we have
$$
(\varepsilon_Y\otimes\varepsilon_X)\circ
(!(e_X\otimes id)\otimes!(id\otimes e_Y))\circ d_{!X\otimes !Y}\circ
m_{!X,!Y}\circ(\delta_X\otimes\delta_Y)
~=~
c_{!Y,!X}^{-1}
~=~
c_{!X,!Y}.
~~~~
\square
$$

\end{document}